# Observation of topological Anderson Chern insulator phase in MnBi$_4$Te$_7$ monolayer

Anqi Wang[1,4,*], Bo Yin[1,4,*], Zikang Su[1,4,*], Shangjie Tian[7], Guoan Li[1,4], Xiaofan Shi[1,4], Xiao Deng[1,4], Yupeng Li[1], Zhiyuan Zhang[1,4], Xingchen Guo[1,4], Qinghua Zhang[1,8], Lin Gu[5], Xingjiang Zhou[1,4,9], Bingbing Tong[1], Peiling Li[1], Zhaozheng Lyu[1], Guangtong Liu[1,9], Fanming Qu[1,9], Ziwei Dou[1], Yuan Huang[3,†], Hechang Lei[2,6,†], Hongming Weng[1,4], Zhong Fang[1,4], Quansheng Wu[1,4,†], Li Lu[1,4,9,†], Jie Shen[1,9,†]

[1]Beijing National Laboratory for Condensed Matter Physics, Institute of Physics, Chinese Academy of Sciences, Beijing 100190, China

[2]School of Physics and Beijing Key Laboratory of Optoelectronic Functional Materials & Micro-Nano Devices, Renmin University of China, Beijing 100872, China

[3]School of Integrated Circuits and Electronics, Beijing Institute of Technology, Beijing 100081, China

[4]School of Physical Sciences, University of Chinese Academy of Sciences, Beijing 100049, China

[5]Beijing National Center for Electron Microscopy and Laboratory of Advanced Materials, Department of Materials Science and Engineering, Tsinghua University, Beijing 100084, China

[6]Key Laboratory of Quantum State Construction and Manipulation (Ministry of Education), Renmin University of China, Beijing 100872, China

[7]Anhui Key Laboratory of Magnetic Functional Materials and Devices, School of Materials Science and Engineering, Anhui University, Hefei 230601, China

[8]Yangtze River Delta Physics Research Center Co. Ltd, Liyang 213300, China

[9]Songshan Lake Materials Laboratory, Dongguan 523808, China

*These authors contributed equally to this work

†Corresponding author. Email: yhuang@bit.edu.cn (Y.H.), hlei@ruc.edu.cn (H.L.), quansheng.wu@iphy.ac.cn (Q.W.), lilu@iphy.ac.cn (L.L.), shenjie@iphy.ac.cn (J.S.)

**Abstract**

The correlation of topology and disorder has attracted great intention due to appropriate disorder could induce the phase transition between trivial and nontrivial topological states. While it is widely recognized that strong disorder can produce rich phase diagrams in topological nontrivial states, moderate disorder has been proposed to induce transitions into topologically nontrivial phases counter-intuitively, leading to the concept of topological Anderson insulators. This phenomenon has been theoretically explored and simulated in various systems, yet experimental realization in solid state systems has remained elusive due to challenges in controlling disorder. Here, we report the experimental observation of Chern insulator state signed by the coexistence of quantized Hall plateau and zero longitudinal resistance in monolayer $MnBi_4Te_7$ Hall bar device, which originally hosts a trivial insulating state with Chern number $C = 0$ in clean limit. We demonstrate that the observed trivial to nontrivial transition in this monolayer device can be attributed to disorder, evidenced by universal conductance fluctuations. Our findings substantiate the existence of a long-sought topological Anderson Chern insulator in real materials, a unique variant of the topological Anderson insulator characterized by broken time-reversal-symmetry.

**Main Text**

Topological quantum material with topologically nontrivial band structure and quantized bulk topological invariants is a major subject in condensed matter physics[1,2]. In conventional wisdom, topological phases in materials are considered robust against weak disorders but will exhibit numerous exotic properties, as well as be inhibited and transition to the trivial phase under strong disorders. Conversely, recent studies have revealed an astonishing reverse transition[3–5]: moderate disorder can trigger the system transition to a topologically nontrivial phase - so-called topological Anderson insulator (TAI) phase, even though the system is topologically trivial in the clean limit. This discovery has sparked much research interest, not only for its brand-new mechanism to realize topological phases beyond the energy band theory but also for its potential to enlarge the parameter space (e.g., chemical potential, temperature, etc.) where topological phases survival in practical application. Due to the attractive interplay and the rich influences between the phase transition and disorder, TAI phase gradually becomes a hallmark for the quantum simulation by several systems, e.g. the cold atoms, photonic crystals, electric circuits etc.[6–10] However, it has never been experimentally observed in real material because the complexity in modulating materials quality and electrostatic atmosphere, or more precisely the degree for the disorder on the basis of required band structure.

Introducing long-range magnetic order in a topological insulator can break the time-reversal-symmetry, leading to the quantum anomalous Hall (QAH) effect with Chiral edge states. This Chern insulator phase without Landau level is attractive not only for its potential application in low-dissipative topological electronic devices that operate without external magnetic field, but also for its capacity in constructing Majorana zero mode for topological quantum computation[11,12]. The QAH effect was first observed in magnetically doped topological insulator[13]. Unfortunately, the inhomogeneity of magnetic dopants limits the material quality and therefore the temperature at which the QAH effect appears. Recently, the QAH effect has been realized in other material systems - the intrinsic magnetic topological insulator $MnBi_2Te_4(Bi_2Te_3)_n$ family[14–16], moiré materials with orbital ferromagnetism[17–19], and spin-orbit proximitized graphene[20,21]. These spark the lights for achieving QAH effect at higher temperature. Notably, the fractional QAH effect have also been realized in moiré materials[22,23], providing a new research platform for studying topology, electron correlations, and topological quantum many-body physics. Furthermore, analogous to the TAI phase, the Chern insulator phase (or the QAH phase) can also be induced by disorder, as predicted theoretically[24] - thus termed as the 'topological Anderson Chern insulator (TACI)' phase.

In this article, we report the observation of disorder-induced TACI phase with non-zero Chern number in the $MnBi_4Te_7$ monolayer device. This monolayer was originally predicted to host trivial insulating states with Chern number $C = 0$ in the clean limit[25]. However, when electrostatically doped from hole($p$) to electron($n$) dominant carriers, it displays a gate-tunable quantum phase transition with universal scaling behavior from a $C = 0$ trivial insulating phase to a $C = 1$ Chern insulator/QAH phase. The Hall resistance $R_{yx}$ quantizes to h/e$^2$, and the longitudinal resistance $R_{xx}$ drops to 0 at a finite magnetic field without the appearance of Landau levels, confirming the Chern insulator phase. Meanwhile, universal conductance fluctuations (UCFs) in magneto-transport measurements appearing in the entire magnetic field range demonstrate the existence of disorders, as well as indicate that it is the origin for the emergence of the Chern insulator phase, in the MnBi$_4$Te$_7$ monolayer device. Moreover, the temperature evolution of magneto-transport reveals the Chern insulator phase with a gap $\Delta E$ ~101 μeV at $\mu_0 H = 12$ T. By comparing these experimental observations with theoretical calculations, we attribute the observed Chern insulator phase to a disorder-induced phase - that is TACI phase - in our device.

**MnBi$_4$Te$_7$ monolayer device**

MnBi$_4$Te$_7$ is a layered van der Waals (vdW) material belonging to the MnBi$_2$Te$_4$(Bi$_2$Te$_3$)$_n$ family which is predicted to be an intrinsic magnetic topological insulator[26–30]. It can be identified as MnBi$_2$Te$_4$ septuple layers (SLs, represented by "*A*") separated by Bi$_2$Te$_3$ quintuple layers (QLs, represented by "*B*") (see the crystal structure in Fig. 1a). Similar to MnBi$_2$Te$_4$, MnBi$_4$Te$_7$ exhibits an A-type antiferromagnetic (AFM) order structure with intralayer ferromagnetic (FM) ordering and interlayer AFM ordering. The QAH effect, axion insulator phase, and high-Chern number Chern insulator state have already been confirmed in MnBi$_2$Te$_4$ devices[14–16,31]. Promisingly, MnBi$_2$Te$_4$(Bi$_2$Te$_3$)$_n$ with higher *n* is supposed to modify the interlayer exchanging coupling (IEC)[32,33]. However, it also induces more difficulty with larger *n* in exfoliating the bulk material into large, flat and uniform thin layers. In this case, MnBi$_4$Te$_7$ might be applicable to achieve a balance between enhancing IEC and overcoming the exfoliation difficulties associated with larger *n* values. The properties of topology and magnetism in MnBi$_4$Te$_7$ have been studied widely in previous works[32–42], but there is still a lack of reports on QAH effect.

Interestingly, according to theoretical calculation, the $MnBi_2Te_4(Bi_2Te_3)_n$ monolayer with a single $MnBi_2Te_4$ SL does not host a non-zero Chern number[25,28–30], which usually appears in multiple-layer devices. Instead, the single $MnBi_2Te_4$ SL is a normal insulator, and *AB* or *BAB* stacking configurations ($MnBi_4Te_7$ monolayer) exhibit a Chern number $C = 0$ state, which might be a time-reversal-symmetry-broken (*T*-broken) quantum spin Hall (QSH) state with spin Chern number $|C_s| = 2$[25]. The absence of a Chern insulator phase in $MnBi_2Te_4$ and $MnBi_4Te_7$ monolayer is also confirmed by subsequent experiment studies where the normal insulator behavior is observed[42,43]. It is remarkable, however, the calculated $C = 0$ phase in *AB* or *BAB* configurations is located near the phase boundary between $C = 0$ and $C = 1$[25]. By tuning the parameters, such as introducing disorder into the system, it is quite promising to realize a phase transition from the $C = 0$ phase to a Chern insulator phase - the above-mentioned TACI. We also notice, by proper exfoliation, $MnBi_4Te_7$ monolayer could contain tunable degree of disorder while still keep good quality for Hall bar measurement, and therefore might be an ideal platform to investigate TACI physics.

The temperature (*T*)-dependent resistance of the $MnBi_4Te_7$ bulk sample used in our experiment (gray curve in Fig. 1d) shows a sharp resistance peak at Néel temperature $T_N \sim 13$ K, consistent with previous reports[32–36]. Benefiting from the vdW nature of $MnBi_4Te_7$ layers, we can transfer $MnBi_4Te_7$ thin films onto $SiO_2/Si^+$ substrates using an $Al_2O_3$-assisted mechanical exfoliation technique[14,44] and subsequently fabricate Hall bar devices (see 'Method' for details). The $Al_2O_3/SiO_2$ layer serves as a back gate dielectric layer, allowing for electrical transport measurements at varied back gate voltages $V_{bg}$ by applying different d.c. voltages to the $Si^+$ layer of the substrate. The Hall bar device (Fig. 1b) based on a $MnBi_4Te_7$ thin film contains a single $MnBi_2Te_4$ SL, confirmed by the cross-sectional scanning transmission electron microscope (STEM) images of the device in Fig. 1c and Extended Data Fig. 1. The STEM images clearly show that the film consists of only one $MnBi_2Te_4$ SL (*A*) on top of one $Bi_2Te_3$ QL (*B*) locating on the amorphous $Al_2O_3$ layer, with slightly residual oxidized

Bi$_2$Te$_3$ QL on the other side. Therefore, it is reasonable to consider this thin film as MnBi$_4$Te$_7$ monolayer.

The blue (red) curve in Fig. 1d displays the temperature-dependent longitudinal (Hall) resistance of the MnBi$_4$Te$_7$ monolayer device at a back gate voltage $V_{bg}$ = 0, with additional data for other gate voltages in Extended Data Figs. 2a, b. The longitudinal resistance rises with the decreasing temperature, indicating a typical semiconducting behavior with a low carrier density. This contrasts with the metallic behavior observed in the bulk sample and confers the advantage of high-efficient modulation via electrostatic gating[14,15,42,43]. Both the longitudinal and Hall curves exhibit a kink at $T_c$ ~ 16 K, with a hysteresis loop appearing in $R_{yx}$ - $\mu_0 H$ curve at the same temperature (Extended Data Fig. 2c). Given the monolayer nature of the device, we attribute $T_c$ to the temperature at which intralayer FM ordering develops. The reason for $T_c$ slightly higher than the Néel temperature $T_N$ of the bulk sample may be the stronger intralayer FM interactions relative to the interlayer AFM interactions[27,34].

**Gate-dependent quantum phase transition from *C* = 0 to *C* = 1 phase**

We first investigate the magneto-transport properties of the device at different back gate voltages ($V_{bg}$) and $T$ = 0.10 K. Figure. 2a demonstrates the representative data with more detailed data provided in Supplementary Figs. 1 - 4. At $V_{bg}$ = -40 V, the Hall resistance ($R_{yx}$) trace shows an obvious positive slope, indicating the Fermi level $E_F$ lies in the valence band with *p*-type carriers dominating. In contrast, for $V_{bg} \geq$ 30 V, $E_F$ lies in the conduction band and the Hall traces exhibit negative slopes due to *n*-type carriers dominance.

As $V_{bg}$ increases from -40 V, we observe a Chern insulator (*C* = 1) phase at gate voltage between ~ 12 V to 18 V (red background in Figs. 2a and b) at high magnetic field. In this region at low magnetic field, both the longitudinal and Hall resistance exhibit hysteresis behavior with a coercive field $\mu_0 H_c$ ~ 0.3 T, followed by a plateau extending to approximately 2 T. With the increasing of magnetic field, $R_{yx}$ rises rapidly and reaches the quantized value of ±h/e$^2$ (h is the Planck constant, and e the charge of an electron) at ~ ± 8 T, accompanied by that $R_{xx}$ drops to 0 concomitantly. (We need to emphasize that, instead of the average value, the minimum (maximum) values of $R_{xx}$ ($|R_{yx}|$) quantize to 0 (1) $h/e^2$, which will be explained in Figs. 3a and b.) This quantization of $R_{yx}$ and vanishing of $R_{xx}$ are hallmarks of Chern insulator state. Nonlocal measurements in Extended Data Fig. 3 demonstrate that the edge modes propagate anticlockwise when magnetization $M > 0$ and clockwise when $M < 0$ in $C = 1$ region. This result is coincident with the predictions from Landauer-Büttiker formula for chiral edge states, further confirming the chiral nature of Chern insulator edges.

It is true that the Landau level quantization at high magnetic field could give rise to the quantum Hall (QH) effect and also a quantized Hall plateau, but it can be ruled out for the following reasons: (1) A strong magnetic field $B$ and high carrier mobility $\mu$ are required for QH effect with filling factor $v = 1$ to satisfy the criterion $\mu B \gg 1$. For the MnBi$_4$Te$_7$ thin film used in this work with $\mu \sim 480$ cm$^2$/V·s (extracted from $R_{yx}$ - and $R_{xx}$ - $\mu_0 H$ curves at $V_{bg} = 16$ V and $T = 25$ K without hysteresis loop), the magnetic field would need to be $\gg 20$ T to generate QH effect with a Hall plateau $R_{yx} = h/e^2$. This value is much larger than $B \sim 8$ T, at which we observe the quantization; (2) No transport evidence of Landau levels (the requirement for QH effect), e.g., quantized Hall plateau with higher filling factors, which is supposed to appear at lower magnetic field, and Shubnikov de Hass oscillations in longitudinal resistance, has been observed in our experiments. (3) The chirality of QH effect is different between $p$-type and $n$-type dominant carriers. However, only one chirality is observed in our experiment when increasing the back gate voltage from - 40 V ($p$-type carriers) to 60 V ($n$-type carriers). The observed chirality in local and nonlocal magneto-transport is also consistent with that of the QAH effect reported in previous studies on MnBi$_2$Te$_4$(Bi$_2$Te$_3$)$_n$[14–16,31]. The above reasons clearly confirm that the observed quantized Hall plateau in this work originates from Chern insulator state without Landau level.

It is worth mentioning that both $R_{xx}$ - and $R_{yx}$ - $\mu_0 H$ curves exhibit apparent resistance fluctuations across the entire magnetic field sweep range. They are UCFs, caused by quantum coherent interference of diffusive paths in disordered mesoscopic samples, instead of random noise. This can be confirmed by the repeatable patterns and the reduction in fluctuation amplitude with increasing temperature (Figs. 3c, d and Extended Data Figs. 4, 5) - the constant disorder configuration contributes to the reproducible quasiperiodic oscillation pattern and the decreasing phase coherence length leads to the decreasing oscillation amplitude with increasing temperature. Using the autocorrelation function method[45,46], we can obtain the phase coherence length $L_\varphi \sim T^{-1/2}$ (Extended Data Fig. 4; see 'Method' for details), indicating that electron-electron interaction in two-dimensions dominate the dephasing mechanism in our sample[45,47].

The clear correlation between the UCFs and Chern insulator state appears on the high magnetic field regions of $R_{xx}$ - and $R_{yx}$ - $\mu_0 H$ curves in Fig. 3a. We can find that the minimum (maximum) of the $R_{xx}$ ($|R_{yx}|$) curve is well quantized to 0 (1) $h/e^2$ while the rest is deviated from this quantized value. Furthermore, the fluctuations in $R_{xx}$ and $R_{yx}$ are in-phase, reaching or deviating from quantized value at the same magnetic field. This correlation is not coincidental but arises naturally from the transport involving chiral edge states in parallel with bulk UCF state (Fig. 3b): The disorders in MnBi$_4$Te$_7$ thin film lead to scattering and diffusive transport in the bulk, coexisting with the chiral edges. At ultra-low temperature, quantum coherent interference occurs, resulting in UCFs. When the interference is destructive, bulk transport channels are suppressed, leaving the dissipationless chiral edge channel which contributes to quantized Hall resistance and zero longitudinal resistance (as shown in left panel in Fig. 3b). Otherwise, the participation of dissipative bulk channels reduces the Hall resistance from $h/e^2$ and renders the non-zero longitudinal resistance (as shown in right panel in Fig. 3b). Note that the $R_{xx}$ and $R_{yx}$ data shown here are anti-symmetrized and symmetrized, respectively (see 'Method' for details). Thus, the observed in-phase fluctuations are intrinsic property of $R_{xx}$ and $R_{yx}$, instead of artifacts of mixing between longitudinal and Hall components. The mesoscopic quantum fluctuations were also observed in submicron size Cr-(Bi,Sb)$_2$Te$_3$ QAH effect devices but restricted to the regions near the coercive field, due to resonant tunnelling between magnetic domains[48]. Differently, the UCFs in our work span the entire magnetic field range and are driven by the disorders in the bulk state of the film.

This correlation between the UCFs and Chern insulator state supports that the disorder plays a crucial role in the emergence of Chern insulator state in this monolayer device, which was originally considered topologically trivial based on both theory and experiments[25,28–30,42,43]. The physics behind this correlation will be discussed in theoretical calculation part later.

Considering the above-mentioned Chern insulator with UCFs scenario, we extract the minimum (maximum) values of $R_{xx}$ ($R_{yx}$) from the $\mu_0 H$-dependent curves in the Chern insulator region and label them as 'x' in Fig. 2b. These values ($R_{xx} < 0.02$ h/e$^2$, $R_{yx} \sim 0.99$ h/e$^2$) are closer to 0 (for $R_{xx}$) and h/e$^2$ (for $R_{yx}$) compared to the $R_{xx}$ ($R_{yx}$) - $V_{bg}$ curves scanning at $\mu_0 H = -12$ T, as they represent the pure edge state of Chern insulator with suppressed bulk conducting state.

Apart from the Chern insulator ($C = 1$) region, we also observe a $C = 0$ region at $V_{bg} = 0$ V (green background in Figs. 2a and b). In this region, $R_{xx}$ increases rapidly with increasing magnetic field, while $R_{yx}$ approaches nearly zero at high magnetic field. This zero $R_{yx}$ feature is more evident in the $R_{yx}$ - and $R_{xx}$ - $V_{bg}$ curves at $\mu_0 H = -12$ T in Fig. 2b (highlighted by green background). Notably, in this region, the longitudinal resistance shows an increasing peak with decreasing temperature, meaning an $C = 0$ insulating state with a band gap. However, the Hall resistance exhibits a zero plateau between ~ -5 V and 6 V (original curves without symmetrization/anti-symmetrization are shown in Supplementary Fig. 5). This behavior is quite different from that of normal insulators which are expected to show diverging large Hall resistance due to the vanish of carriers. The scenario of the coexistence of electron and hole carriers which could result in zero Hall resistance is unlikely here, because it can only appear with a certain parameter and produce a zero Hall 'point' instead of a zero Hall 'plateau' as shown in Fig. 2b[49–52]. Axion insulator phase resulting from the cancel of opposite Chern numbers in AFM topological insulator is another scenario for zero

Hall resistance[15,53,54]. But it should manifest in the AFM phase at near-zero magnetic fields instead of FM phase under high magnetic fields. Considering previous theory prediction for MnBi$_4$Te$_7$ monolayer, this $C = 0$ region may be attributed to the $T$-broken QSH state[25]. But there is still a lack of evidence to verify the nature of this $C = 0$ state in our experiment.

Thus, there is a quantum phase transition from the trivial $C = 0$ phase to the non-trivial Chern insulator ($C = 1$) phase by electrostatic gating. This gate-dependent quantum phase transition is analogous to the plateau/insulator-to-plateau transition in quantum Hall (QH) system[55–58] and quantum phase transition in QAH system[59]. When we choose $(V_{bg} - V_{bg}^T)/T^\kappa$ as $x$-axis, $R_{xx}(V_{bg})$ curves at different $T$ below 0.50 K collapse as one curve (Fig. 2c), demonstrate the single-parameter scaling behavior in quantum phase transition from $C = 0$ to $C = 1$. This quantum phase transition and universal scaling behavior reveals a topological trivial gap and a topological nontrivial (Chern insulator) gap coexist in our system but locate at different back gate region.

**Evolution of Chern insulator phase**

To further probe the temperature evolution of Chern insulator state in our device, we measure the magnetic-field-dependent $R_{yx}$ and $R_{xx}$ from 0.05 K to 25 K at $V_{bg} = 16$ V (Figs. 3c and d, with a 0.5 h/e$^2$ offset for curves in different $T$) and summarize the $R_{yx}$ and $R_{xx}$ values at -12 T in Fig. 3e. The 'x' labels in Fig. 3e demonstrate $R_{yx}$ extracted from maximum values in $R_{yx}$ - $\mu_0 H$ curves and represent the pure edge state of Chern insulator without bulk state according to the previous analysis in Fig. 3b. At temperatures below ~ 0.20 K, the Hall resistance is well quantized and the longitudinal resistance vanishes. As the temperature rises above 0.2 K, $R_{yx}$ starts to deviate from the quantized value, and $R_{xx}$ exhibits thermally activated behavior[14]. Fitting $R_{xx}$ to the Arrhenius relation $R_{xx}$ ~ exp(-$\Delta E/2k_B T$) (k$_B$ is the Boltzmann constant and $\Delta E$ the energy gap) gives a gap $\Delta E$ ~101 μeV at $\mu_0 H = 12$ T in the Chern insulator phase.

The $R_{xx}$ - $\mu_0 H$ curves at different temperatures intersect at a crossing point with $R_{xx}$ ~ 0.5 h/e$^2$ and $H_T$ ~ 2.5 T (Fig. 3f), indicating a magnetic-field-dependent quantum phase transition from an insulating phase at low field to the Chern-insulator phase at high field[60–62]. The insulating phase is characterized by an increase in $R_{xx}$ with decreasing temperature while the Chern insulator phase shows the opposite behavior, as demonstrated by the temperature dependence of $R_{xx}$ at different magnetic fields in Fig. 3h. There are possibly multiple magnetic domains, which are induced by oxidation/device fabrication, appearing at low magnetic field. These domains lead to the scattering and tunneling between the chiral edges at domain boundaries. As a result, the insulating phase emerges at low field and will transition to Chern insulator phase when domain flips to form a uniform FM state at high field.

This magnetic-field-dependent quantum phase transition can also be scaled using the single-parameter scaling function $(H-H_T)/T^\kappa$ as shown in Fig. 3g. The $R_{xx}$ v.s. $(H-H_T)/T^\kappa$ curves at different $T$ collapse into one, demonstrating the universal scaling behavior below $T = 0.80$ K. This universal scaling behavior also appears in quantum phase transition of QH and QAH systems, where the critical exponent $\kappa$ depends on the nature or range of the disorder and therefore sometimes varies between different samples or even between different transitions within a single sample[63]. The different values of $\kappa$ in gate- and magnetic-field-dependent quantum phase transition in our sample is still unresolved and need further exploration.

**Theoretical calculation**

To further elucidate the experimental results, we conduct theoretical calculations using non-equilibrium Green's function (NEGF) quantum transport in a six-terminal Hall bar configuration alongside parameters derived from first-principles calculations. These studies investigate the coexistence of the $C = 0$ state and the TACI state as a function of the Fermi energy.

Initially, we employ first-principles calculations and a low-energy effective k·p model to fit the band structure of $MnBi_4Te_7$ (see Fig. 4a). The parameters are selected based on previous studies of $Bi_2Se_3$, $Bi_2Te_3$, and $MnBi_2Te_4$ systems[30,64]. The band structures of a ribbon shown in Fig. 4b confirm the $T$-broken QSH state in the clean limit, consistent with earlier theoretical predictions[25]. To simulate the disorder effect, we introduce an on-site random magnetic disorder term $H_D = V \cdot \sigma_z \otimes \tau_z$, where $\sigma$ and $\tau$ acting on spin and orbital degrees of freedom, respectively. Disorder's onsite energy $V$ is uniformly distributed in the range [-$W$/2, $W$/2] with disorder strength $W$. To mitigate random fluctuations, we average over multiple disorder configurations.

Subsequently, using the NEGF formalism within the Landauer-Büttiker framework we calculate $R_{xx}$ and $R_{yx}$ as functions of the Fermi energy ($E_F$). As the disorder strength $W$ increases from the clean limit, the $T$-broken QSH state is destroyed, evolving into the TACI state. Further increasing $W$ induces an Anderson transition, leading the system into an Anderson insulator state. For TACI state at $W$ = 8.5, detailed results of $R_{xx}$ and $R_{yx}$ versus $E_F$ are presented in Fig. 4c and Fig. 4d. Sample sizes range from 100*100 to 500*500, with 800 for size 100 and size 200, 100 for size 500 averaging to simulate the thermodynamics limit.

In the valence band region -0.7 eV ~ -0.4 eV of $E_F$, $R_{xx}$ fluctuates around 0.5 h/e$^2$, while $R_{yx}$ vanishes with significant fluctuations and matches the green region Fig. 2b. In particular, as the sample size increases, $R_{xx}$ grows in magnitude, suggesting a possibly peak in the thermodynamics limit, consistent with the peak observed in the green region of Fig. 2b. Transmission coefficient analysis (right panel in Fig. 4f) indicates the presence of only two edge channels, with nonzero transmission coefficients existing solely between nearest leads but reduced below 1 due to impurity scattering. Consequently, $R_{xx}$ exceeds 0.5 h/e$^2$ in the $C$ = 0 state, while $R_{yx}$ remains negligible and highly fluctuating, as observed in Fig. 2b. In the gap region -0.02 eV ~ 0.06 eV of $E_F$, $R_{xx}$ decreases and eventually vanishes, while $R_{yx}$ raises to form a plateau, aligning with the red region in Fig. 2b. As the sample size increases, $R_{yx}$ approaches a nearly quantized value of ~ 0.97 h/e$^2$, and $R_{xx}$ decreases to ~ 0.08 h/e$^2$, consistent with the quantization observed at $V_{bg}$ ~ 16 V in Fig. 2b. Transmission coefficient analysis (left panel in Fig. 4f) reveals the existence of a single quantized edge channel between nearest leads, confirming the TACI state.

Figure 4e presents the phase diagram of $R_{yx}$ as a function of disorder strength $W$ and Fermi energy $E_F$. Due to the small gap in the real material, scattering between edge states occurs readily, necessitating sufficiently large sample sizes in numerical simulations to minimize scattering effects. Although the plateau of $R_{yx}$ for the TACI region is approximately 0.9 due to computational limitations, it is sufficient to distinguish between different phases and approaches quantization with larger sample sizes, as supported by Fig. 4d. Finally, the phase diagram for TACI indicates that disorder plays a pivotal role in driving topological phase transitions akin to the TAI. While the TACI and TAI phase diagrams share similarities, they differ in the types and forms of phases. A distinctive feature is the coexistence of $C = 0$ state and TACI state versus Fermi energy which is observed in the experiment as well, although whether the $C = 0$ state is exactly $T$-broken QSH state in experiment still needs further exploration. Unlike the conventional Chern insulator state, which exists within the gap in the clean limit, the TACI state experimentally observed and theoretically confirmed can only exist in disordered conditions and vanishes in the clean limit. This finding potentially broadens the parameter space for realizing the QAH effect.

**Discussion**

We note that the reentrant QAH effect in $MnBi_2Te_4$ thin films has been reported during the same period of our research and the researcher termed this phenomenon as 'Chern Anderson insulator (CAI)'[65]. This CAI phase is different from the TACI phase observed in our study. In CAI phase, there is already a QAH phase ($C = 1$) in the charge neutral point of the 5SL $MnBi_2Te_4$ thin film, and the QAH effect reenter in the conducting state due to exchange-induced Berry curvature splitting and disorder-induced Anderson localization. In our TACI phase, the $MnBi_4Te_7$ monolayer is topologically trivial ($C = 0$) without considering the disorders based on both theory and experiments[25,28–30,42,43]. And there is apparently a gate-dependent quantum phase transition with universal scaling behavior from $C = 0$ to $C = 1$ state due to the disorder effect, supporting the demonstration of TACI phase emerging from the topologically trivial phase. Moreover, the disorder-induced in-phase UCFs and the quantized values at the minimal (maximal) of the $R_{xx}$ and $R_{yx}$ - $\mu_0 H$ curves further highlight the correlation between disorder and Chern insulator state in the TACI phase. These disorders comes from native defects in $MnBi_2Te_4(Bi_2Te_3)_n$ family or device exfoliation/fabrication, and are more obvious in monolayer device with higher *n* studied in our research[36,66–70].

We conclude the experimental observation of such a TACI in solid state system suggests that the inclusion of disorder in condensed matter systems can play a pivotal role in facilitating the realization of QAH effect, particularly in expanding the parameter space necessary for its manifestation. Not only would this advance our fundamental understanding of quantum phenomena, but it would also open the door to new electronic devices that leverage the unique properties of materials exhibiting QAH effect.

## Methods

### Single crystal growth

MnBi$_4$Te$_7$ single crystals were grown by self-flux method. Mn piece, Bi lump and Te lump were weighed with the molar ratio of Mn : Bi : Te = 1 : 8 : 13 (MnTe : Bi$_2$Te$_3$ = 1 : 4). The mixture was loaded into a corundum crucible sealed into a quartz tube. The tube was then placed into a furnace and heated to 1100 °C for 20 h to allow sufficient homogenization. After a rapid cooling to 600 °C at 5 °C/h, the mixture was cooled slowly to 585 °C at 0.5 °C/h and kept at this temperature for 2 days. Finally, the single crystals were obtained after centrifuging. The plate-like MnBi$_4$Te$_7$ single crystals can be easily exfoliated.

### Device fabrication

To prevent oxidation, all the fabrication processes, except electron-beam lithography, were carried out in a nitrogen-filled glove box. The MnBi$_4$Te$_7$ thin film was exfoliated and transferred onto a SiO$_2$/Si$^+$ substrate using the Al$_2$O$_3$-assisted mechanical exfoliation technique described in ref. 44 and then spin-coated with PMMA in the glove box. The Hall bar geometry was patterned using electron-beam lithography and moved back into the glove box to develop. Then we deposited the Cr/Au electrodes using a thermal evaporation system installed in the glove box and performed lift-off process without being moved out of the glove box. After this, we repeated the PMMA-coating and electron-beam lithography process once again to expose the Ti/Au pads for wire-bonding. The other parts of the device were protected with PMMA, so we could take the device outside the glove box for wire-bonding and measurements. The MnBi$_4$Te$_7$ thin film was in nitrogen atmosphere or protected with PMMA in the air (to perform electron-beam lithography) during the whole fabrication process.

**Electrical transport measurements**

The electrical transport measurements were performed in a dilution refrigerator (Oxford instruments, ~ 10 mK, 12 T). The back gate voltage $V_{bg}$ was applied using a Keithley 2400. Standard lock-in measurements were taken with a small a.c. excitation current (1.5 nA) to avoid the breakdown of the Chern insulator state.

**Symmetrization and anti-symmetrization procedures**

We employ the standard symmetrization and anti-symmetrization procedures to separate the $R_{xx}$ and $R_{yx}$ components in the measurement. In magnetic field sweeping process, the symmetrization and anti-symmetrization procedures follow:

$$R_{xx}^{\uparrow}(\mu_0 H) = [R_{xx,\text{raw}}^{\uparrow}(\mu_0 H) + R_{xx,\text{raw}}^{\downarrow}(-\mu_0 H)]/2,$$

$$R_{xx}^{\downarrow}(\mu_0 H) = [R_{xx,\text{raw}}^{\downarrow}(\mu_0 H) + R_{xx,\text{raw}}^{\uparrow}(-\mu_0 H)]/2,$$

$$R_{yx}^{\uparrow}(\mu_0 H) = [R_{yx,\text{raw}}^{\uparrow}(\mu_0 H) - R_{yx,\text{raw}}^{\downarrow}(-\mu_0 H)]/2,$$

$$R_{yx}^{\downarrow}(\mu_0 H) = [R_{yx,\text{raw}}^{\downarrow}(\mu_0 H) - R_{yx,\text{raw}}^{\uparrow}(-\mu_0 H)]/2,$$

where 'raw' in subscription means raw data, and '↑'/'↓' in superscription indicate the magnetic field sweeping direction.

In gate voltage sweeping process at $\mu_0 H = \pm 12$ T, the symmetrization and anti-symmetrization procedures follow:

$$R_{xx}^{12\text{T}}(V_{bg}) = R_{xx}^{-12\text{T}}(V_{bg}) = [R_{xx,\text{raw}}^{12\text{T}}(V_{bg}) + R_{xx,\text{raw}}^{-12\text{T}}(V_{bg})]/2,$$

$$R_{yx}^{12\text{T}}(V_{bg}) = -R_{yx}^{-12\text{T}}(V_{bg}) = [R_{yx,\text{raw}}^{12\text{T}}(V_{bg}) - R_{yx,\text{raw}}^{-12\text{T}}(V_{bg})]/2.$$

Due to the pronounced heating effects during magnetic field sweeps at ultra-low temperatures, we can only use a very low magnetic field sweep rate for measurements at $T = 0.05$ K. In this condition, we performed bi-directional sweeps within the magnetic field range of ±1 T, but only sweep one direction for $|\mu_0 H| > 1$ T in $R_{xx}/R_{yx}$ - $\mu_0 H$ curves at $T = 0.05$ K and $V_{bg} = 16$ V, owing to the limited measurement time. This adjustment doesn't affect the results, since there is no hysteresis loop in the magnetic field range beyond ±1 T. All other $R_{xx}/R_{yx}$ - $\mu_0 H$ curves are measured with both two directions within the magnetic field range of ±12 T.

**Extract phase coherence length from UCFs**

To extract phase coherence length $L_\varphi$ v.s. $T$ from UCFs, we continuously measure the $R_{xx}$ v.s. $\mu_0 H$ at different $T$ in one thermal cycle (Extended Data Fig. 4a) and subtract the resistance background to extract the UCFs (Extended Data Fig. 4b). The UCFs are quasiperiodic with a characteristic correlation field scale $\Delta B_c$, which can be determined from the half maximum of the autocorrelation function of the fluctuations as shown in Extended Data Fig. 4c[45,46]. The phase coherence length can be extracted from $\Delta B_c$ using the relation:

$$\Delta B_c = C\Phi_0/WL_\varphi ,$$

Where $\Phi_0$ = h/e is the magnetic flux quantum, $W$ is the width of the sample, and $C \sim 0.95$ for $L_\varphi \gg L_T$ (thermal length). The extracted phase coherence length $L_\varphi$ is demonstrated in Extended Data Fig. 4d.

**Hysteretic behavior in gate sweeps**

The $R_{xx}/R_{yx}$ v.s. $V_{bg}$ curves exhibit hysteresis behavior due to the charging effect of the $Al_2O_3/SiO_2$ back gate dielectric layer. As shown in Supplementary Fig. 5a, when the gate voltage is swept in the same direction, the curves are almost coincident. However, when swept in opposite direction, a gate voltage shift of ~ 5 V is observed. To perform the symmetrization/anti-symmetrization procedures for $R_{xx}/R_{yx}$ - $V_{bg}$ curves, we ensure the gate voltages swept at $\mu_0 H = \pm 12$ T are in the same direction at same $T$, thereby avoiding the hysteresis effect. For symmetrized/anti-symmetrized $R_{xx}/R_{yx}$ - $V_{bg}$ curves at different $T$, we use the similar method as used in ref. 59 to correct the gate voltage hysteresis: by applying a ~ 5 V gate voltage shift in curves with different sweeping direction and make the $R_{xx}$ - $V_{bg}$ curves crossover at one point (the phase transition point $V_{bg}^T$). Supplementary

Figure 5b demonstrates the original $R_{xx}/R_{yx}$ - $V_{bg}$ curves sweeping at $\mu_0 H = \pm 12$ T and different $T$ and Supplementary Fig. 5c gives the specific gate voltage shifts applied to each curve to correct the hysteretic behavior. The $R_{xx}/R_{yx}$ - $V_{bg}$ curves after gate voltage correction are shown in Fig. 2b.

**Theoretical calculation**

<u>Ab initio calculations</u>

We employed the Vienna ab initio simulation package (VASP)[71] to simulate electronic properties of paramagnetic monolayer MnBi$_4$Te$_7$ in the framework of density functional theory (DFT)[72,73]. The calculation with the Perdew–Burke–Ernzerhof functional (PBE)[74,75] in the generalized gradient approximation. A cutoff energy of 500 eV and a k-mesh of 8 × 8 × 1 are adopted. The calculations are performed with the spin-orbit coupling and $U = 5$ eV for $d$ orbitals of Mn[25,76]. The electronic structures of monolayer MnBi$_4$Te$_7$ are shown in Fig. 4a.

<u>Low-energy effective model</u>

The crystal structure of monolayer MnBi$_4$Te$_7$ is trigonal with the space group $P\bar{3}m1$ (No. 156) and has the same point group as that of monolayer MnBi$_2$Te$_4$ with space group $R\bar{3}m$ (No. 160)[77].

To capture the topological properties, a simple effective Hamiltonian can be written down near the Γ point to characterize the long-wavelength properties of the monolayer MnBi$_4$Te$_7$ system at low energy. Due to the same point group generators as MnBi$_2$Te$_4$, they share an identical k·p effective Hamiltonian[24,30,78]:

In the basis of ($|P1_z^+,\uparrow\rangle$, $|P2_z^-,\uparrow\rangle$, $|P1_z^+,\downarrow\rangle$, $|P2_z^-,\downarrow\rangle$), the effective Hamiltonian at the Γ point is:

$H = H_N + H_{FM}$, where

$$H_N = \varepsilon_0(k) + \begin{pmatrix} M(k) & 0 & 0 & A_1 k_- \\ 0 & -M(k) & A_1 k_- & 0 \\ 0 & A_1 k_+ & M(k) & 0 \\ A_1 k_+ & 0 & 0 & -M(k) \end{pmatrix}$$

and

$$H_{FM} = \begin{pmatrix} M_1(k) & 0 & 0 & A_2 k_- \\ 0 & M_2(k) & -A_2 k_- & 0 \\ 0 & -A_2 k_+ & -M_1(k) & 0 \\ A_2 k_+ & 0 & 0 & -M_2(k) \end{pmatrix}$$ for the correction in the presence of magnetization.

Here $M(k) = M_0 + Bk^2$, $\varepsilon_0(k) = C + Dk^2$, $k_\pm = k_x \pm ik_y$, $M_{1,2}(k) = M_{1,2} + B_{1,2}k^2$, and $\varepsilon_0(k)$ is the particle-hole asymmetric term for real material. By fitting to the DFT calculations, the parameters are $M_0 = 0.095$ eV, $B = -53.49$ eV·Å$^2$, $M_1 = 0.04$ eV, $B_1 = -17.44$ eV·Å$^2$, $M_2 = 0.01$eV, $B_2 = -8.62$ eV·Å$^2$, $C = 0.012$eV, $D = 20.16$ eV·Å$^2$, $A_1 = 2.82$ eV·Å, $A_2 = -1.8$ eV·Å.

We transfer the effective Hamiltonian from the momentum space into the lattice space with the lattice constant a = 4.355 Å. As shown in Fig. 4b, the monolayer MnBi$_4$Te$_7$ system is *T*-broken QSH effect with two nondegenerate Dirac cones (green lines) and is consistent with the previous calculations[25].

Non-equilibrium Green's Function Methods for quantum transport

We calculate transport properties of the six-terminal Hall bar device (see Fig. 4f) by using the Landauer-Büttiker formula[79–81] and the recursive Green's function method[82,83]. The current in the lead *p* with spin index σ is

$$I_{p\sigma} = \frac{e}{\hbar} \sum_{q \neq p} T_{pq}^\sigma (V_{p\sigma} - V_{q\sigma}),$$

where $T_{pq}^\sigma$ stands for the transmission coefficient from lead *p* to lead *q* with spin σ, and $V_{p\sigma}$ is the voltage in the lead *p* with spin index σ.

The transmission coefficient can be obtained from

$$T_{pq} = Tr[\Gamma_p G^r \Gamma_q G^a],$$

where the linewidth function $\Gamma_{p\sigma} = i[\Sigma_{p\sigma}^r - \Sigma_{p\sigma}^a]$, $\Sigma_{p\sigma}^r$ is the retarded self-energy at the lead *p* with spin σ, for simplicity, we choose the same Hamiltonian as the central region to get the retarded self-energy with surface Green's function[84,85].

The retarded Green function of the system can be calculated from

$$G^r = [G^a]^\dagger = (E_F I - H_c - \sum_{p\sigma} \Sigma_{p\sigma}^r)^{-1},$$

where $E_F$ is the Fermi energy and $H_c$ is the Hamiltonian of the system.

To compare with the experimental results, we perform the simulation when the current is applied between lead 1 and lead 4, and measure the voltage between lead 2 and lead 3 to get the longitudinal resistance $R_{xx}$ and the Hall resistance $R_{yx}$, see Fig. 4f,

$$R_{xx} = V_{23}/I_{14},$$

$$R_{yx} = V_{26}/I_{14}.$$

In the simulations, we choose the size of lead is 40% of the central length, leaving 30% on both sides of lead 1 and lead 4.

Next, we consider the on-site random magnetic disorder $H_D = V \cdot \sigma_z \otimes \tau_z$, where σ and τ acting on spin and orbital degrees of freedom, respectively, and $V$ is uniformly distributed in the range [-$W$/2, $W$/2] with disorder strength $W$.

## Data availability

All data needed to evaluate the conclusions in the paper are present in the main text and/or the supplementary information. The authors declare that all of the raw data generated in this study have been deposited in Figshare (Update before publishment).

## Acknowledgments

We are grateful to X. Xie, S. Shen, X. Wan, C. Chen, H. Jiang, Y. Feng, and J. Li for helpful discussions. The work of J.S., L.L., Z.D., F.Q. and G.L. were supported by the National Key Research and Development Program of China (Grant Nos. 2023YFA1607400, 2024YFA1613200), the Beijing Natural Science Foundation (Grant No. JQ23022), the Strategic Priority Research

Program B of Chinese Academy of Sciences (Grant No. XDB33000000), the National Natural Science Foundation of China (Grant Nos. 92065203, 12174430, and 92365302), and the Synergetic Extreme Condition User Facility (SECUF, https://cstr.cn/31123.02.SECUF). The work of other authors were supported by the National Key Research and Development Program of China (Grant Nos. 2019YFA0308000, 2022YFA1403800, 2023YFA1406500, and 2024YFA1408400), the National Natural Science Foundation of China (Grant Nos. 12274436, 12274459, and 12404154), the Beijing Natural Science Foundation (Grant No. Z200005), and the Synergetic Extreme Condition User Facility (SECUF, https://cstr.cn/31123.02.SECUF).

**Author contributions**

J.S. conceived and designed the experiment.

A.W. and Z.S. fabricated devices and performed the transport measurements, with the help of G.L., X.S., X.D., Y.L., Z.Z, and X.G., supervised by B.T., P.L., Z.L., G.L., F.Q., Z.D., L.L., and J.S..

B.Y. performed the theoretical calculations supervised by H.W., Z.F., and Q.W..

S.T. grew bulk $MnBi_4Te_7$ crystals supervised by H.L..

Q.Z., and L.G. performed the STEM measurements.

H.Y. and X.Z. provided supports on thin film exfoliation and devices fabrication.

A.W., B.Y., Q.W., and J.S. analyzed the data and wrote the manuscript, with input from all authors.

**Competing interests**

Authors declare that they have no competing interests.

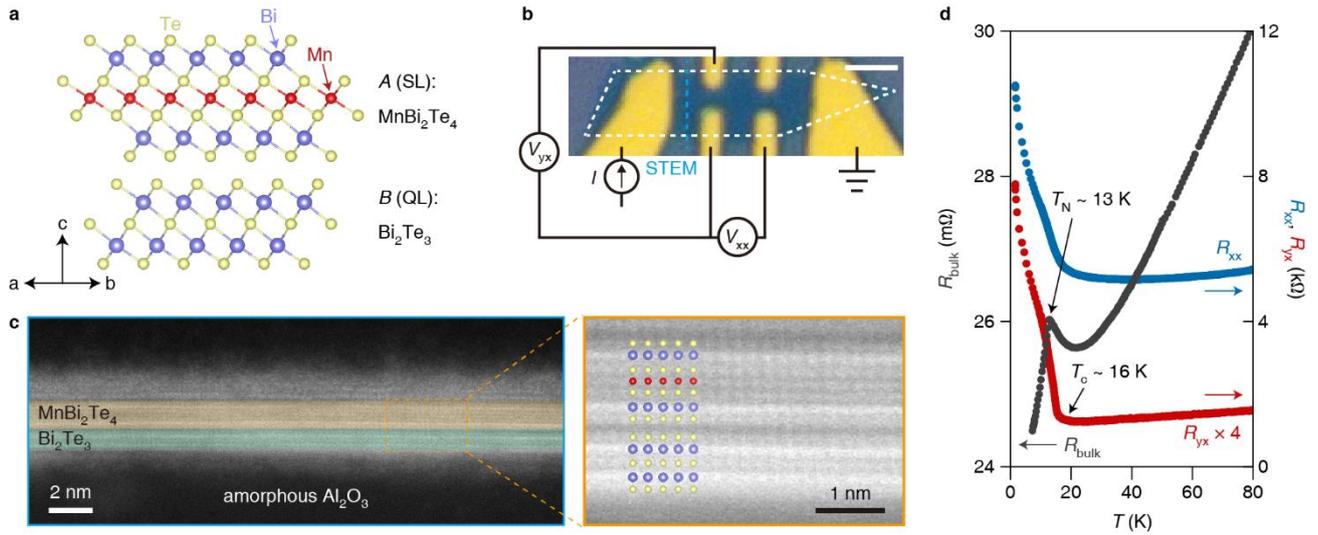

**Fig. 1 MnBi$_4$Te$_7$ monolayer device. a,** Crystal structure of MnBi$_4$Te$_7$. **b,** Optical image of the MnBi$_4$Te$_7$ monolayer device and schematic measurement configuration. The dashed white line highlights the MnBi$_4$Te$_7$ monolayer thin film. Scale bar, 2 μm. **c,** Cross-sectional STEM image of the monolayer device (Scan from the cross-section as blue line shows in **(b)**, the zoom-in image with orange box is filtered to demonstrate the crystal structure.). **d,** Temperature-dependent MnBi$_4$Te$_7$ bulk sample resistance (gray) and longitudinal/Hall (blue/red) resistance of the monolayer device at $V_{bg} = 0$ V.

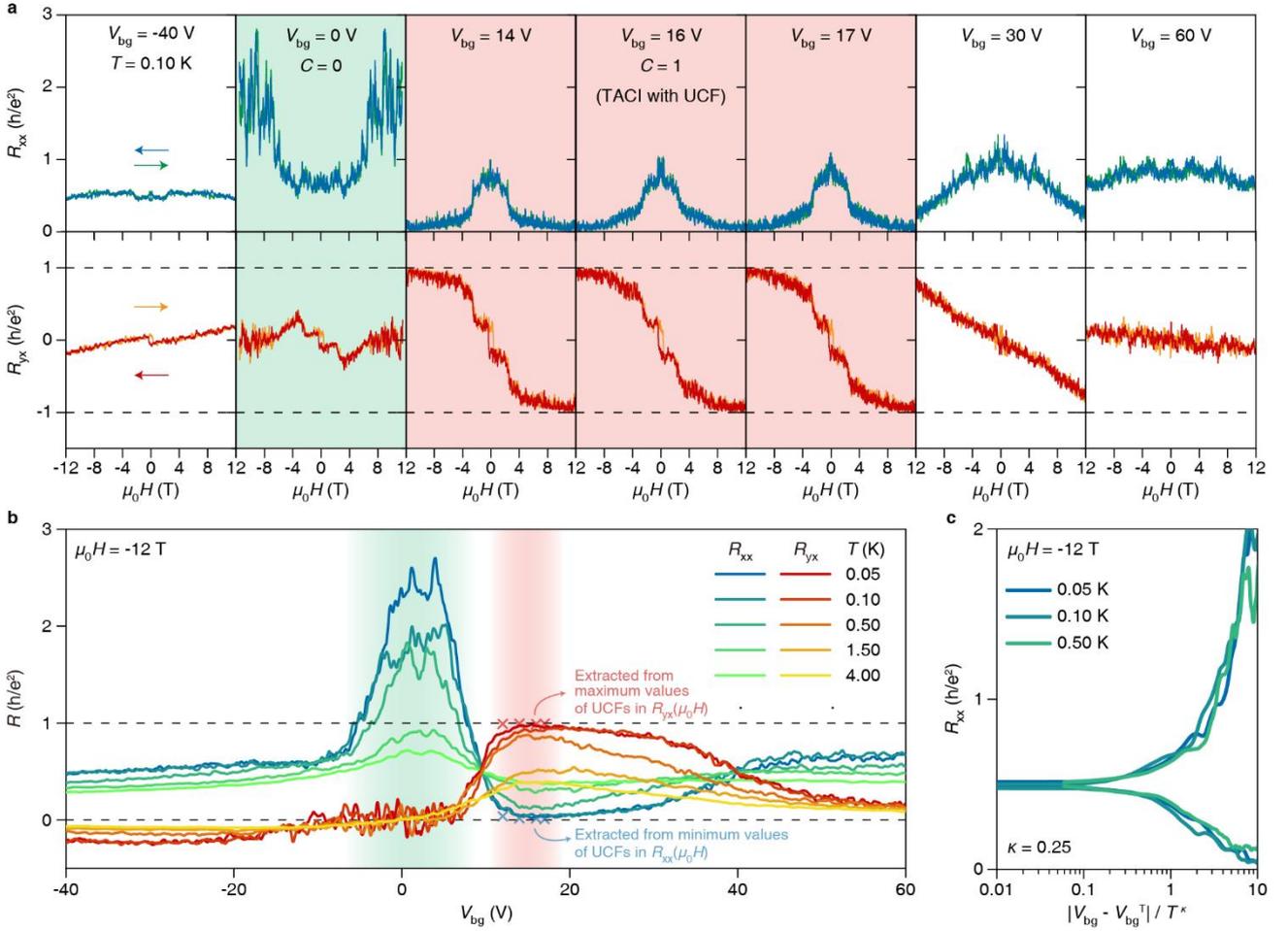

**Fig. 2 Gate-dependent quantum phase transition in monolayer device. a,** $R_{xx}$ - and $R_{yx}$ - $\mu_0 H$ curves at different back gate voltages and $T = 0.10$ K. **b,** $R_{xx}$ and $R_{yx}$ versus back gate voltage at $\mu_0 H$ = -12 T (UCFs are near destructive interference) and different $T$. 'x' labels represent $R_{xx}$ ($R_{yx}$) extracted from minimum (maximum) values of UCFs in $R_{xx}$ ($R_{yx}$) - $\mu_0 H$ curves at $T = 0.10$ K. Green and red background in **(a, b)** represent $C = 0$ and $C = 1$ (TACI) regions, respectively. **c,** $R_{xx}$ - |$V_{bg}$-$V_{bg}^T$|/$T^{0.25}$ at $\mu_0 H$ = -12 T and different $T$ to demonstrate universal scaling behavior in gate-dependent quantum phase transition.

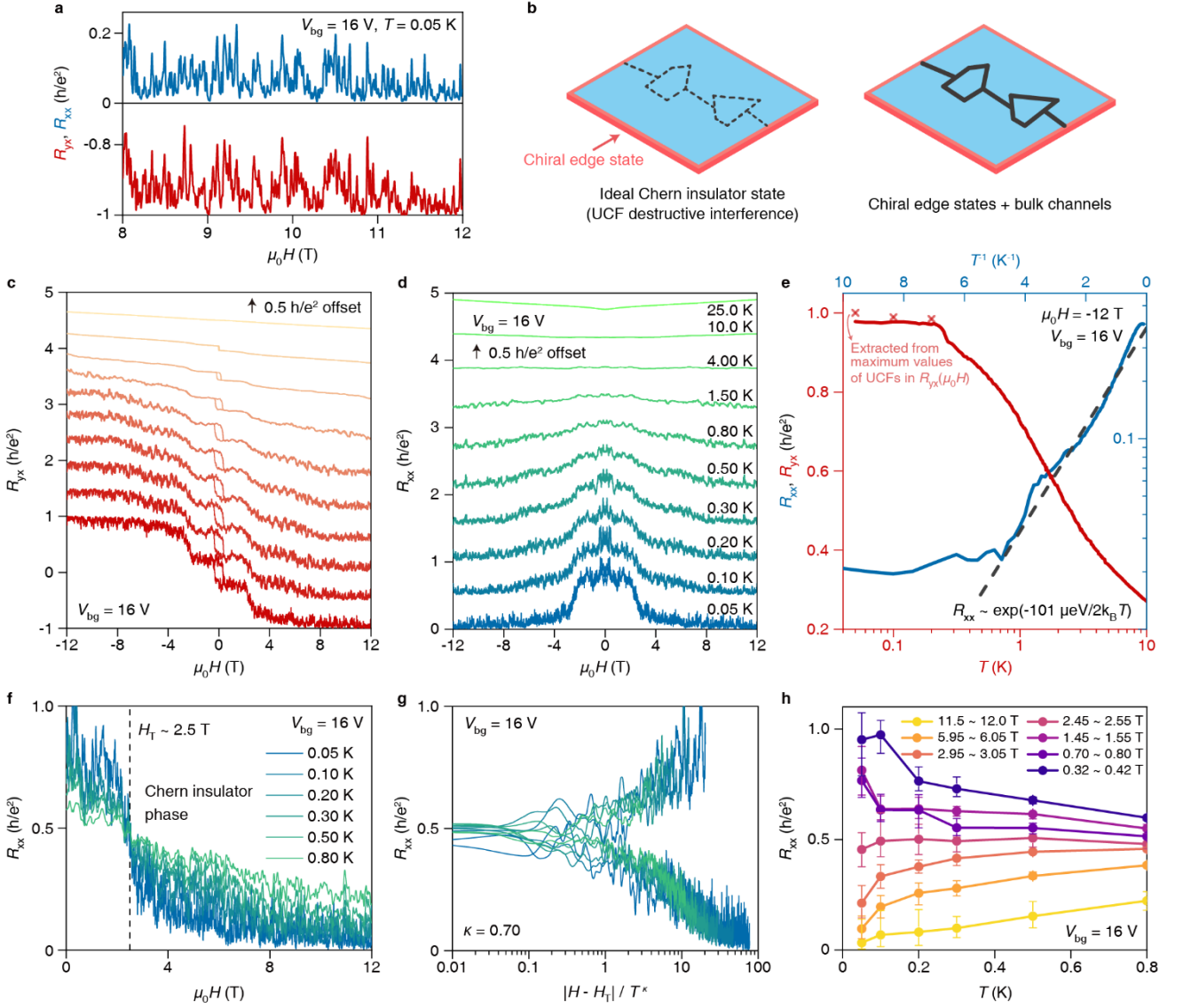

**Fig. 3 Temperature evolution of the TACI phase in monolayer device. a,** Zoom-in plot of $R_{xx}$ - and $R_{yx}$ - $\mu_0H$ curves at $V_{bg}$ = 16 V and $T$ = 0.05 K within the magnetic field range of 8 ~ 12 T. **b,** Schematic of two different Chern insulator scenarios. Blue regions represent Chern insulators, red lines represent chiral edges, gray solid and dashed lines represent bulk conductive and destructive interference channels, respectively. **c, d,** Temperature evolution of $R_{yx}$ **(c)** and $R_{xx}$ **(d)** v.s. $\mu_0H$ curves at $V_{bg}$ = 16 V with 0.5 h/e² offset. **e,** Temperature-dependent $R_{yx}$ and $R_{yx}$ at $\mu_0H$ = - 12 T (UCFs are near destructive interference) and $V_{bg}$ = 16 V. The gray dashed line represents the fitting to the Arrhenius relation. 'x' labels represent $R_{yx}$ values extracted from maximum of UCFs in $R_{yx}$ - $\mu_0H$ curves. **f,** The same data as **(d)** at $T$ below 0.8 K without offset to demonstrate quantum phase

transition. **g,** The same data as **(f)**, but plotted as $R_{xx}$ - $|H-H_T|/T^{0.70}$ to demonstrate universal scaling behavior in magnetic-field-dependent quantum phase transition. **h,** Temperature-dependent $R_{xx}$ at varied magnetic field extracted from **(f)**. The dots represent the mean values of $R_{xx}$ in magnetic field range shown in the legends. The error bars reflect the maximum and minimum values in the same range.

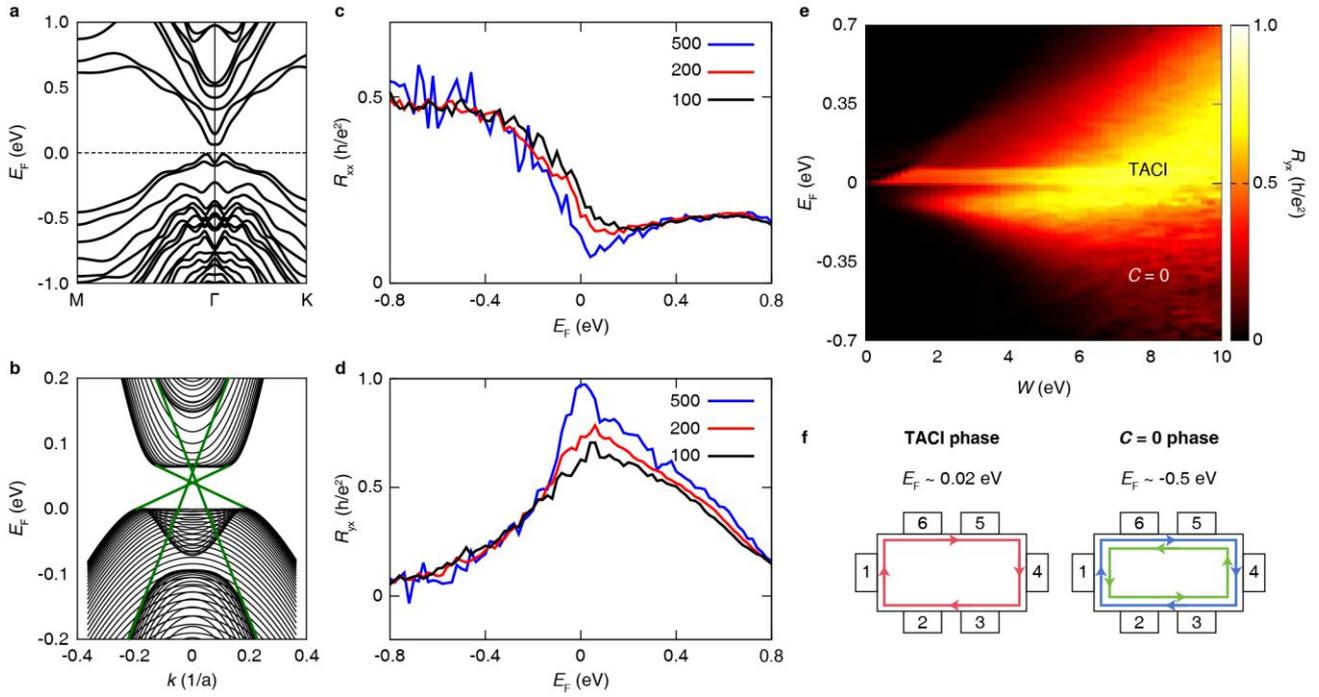

**Fig. 4 First-principles calculations, low-energy effective model, and transport properties of monolayer MnBi$_4$Te$_7$. a,** The electronic structures of monolayer MnBi$_4$Te$_7$. **b,** Band structure of ribbon with open boundary of size 200a. **c, d,** $R_{xx}$ **(c)** and $R_{yx}$ **(d)** versus $E_F$ at $W$ = 8.5 of sample size from 100*100 to 500*500, averaging over 800, 800, 100 samples for size 100, 200, 500. **e,** Phase diagram of $R_{yx}$ versus $W$ and $E_F$, sample size is 200*200, averaging over 50 samples. **f,** The Hall bar and the transmission coefficient analysis at $E_F$ ~ 0.02 eV (TACI phase) and $E_F$ ~ -0.5 eV ($C$ = 0 phase).

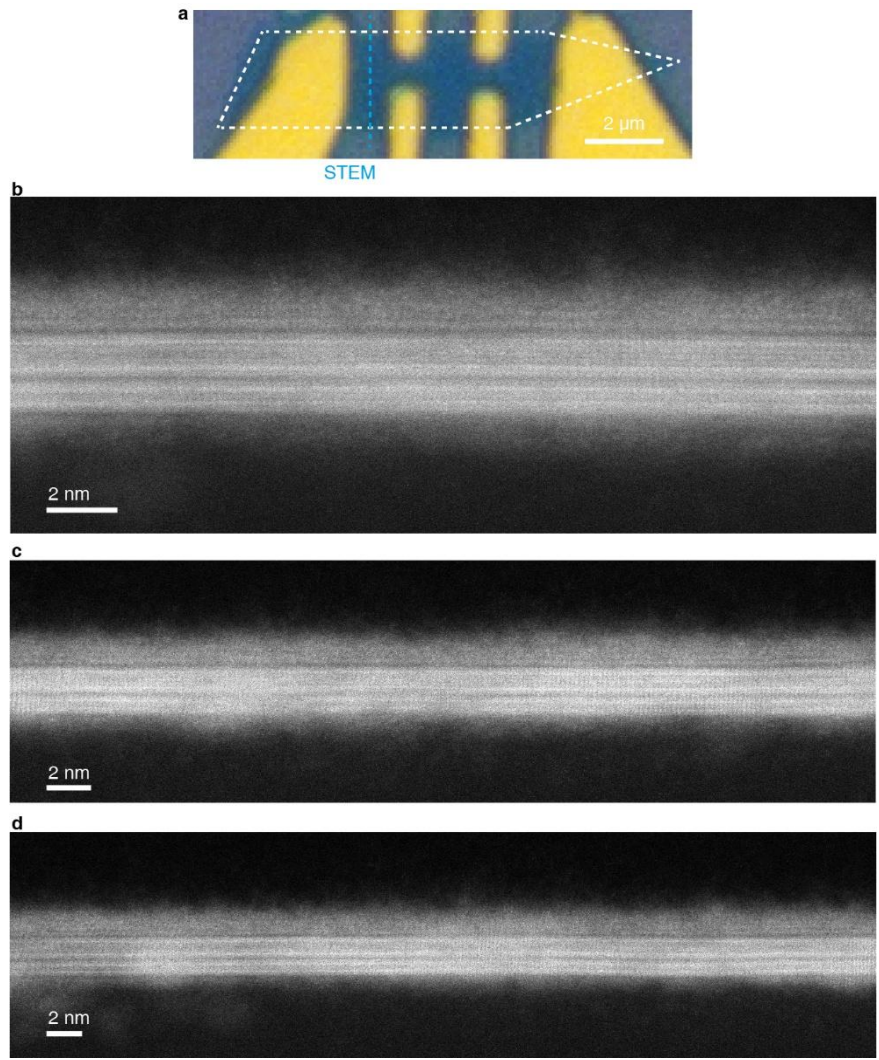

**Extended Data Fig. 1 Cross-sectional STEM images of the monolayer device. a,** Optical image of the MnBi$_4$Te$_7$ monolayer device. The dashed white line highlights the MnBi$_4$Te$_7$ monolayer thin film. **b,** The same image of Fig. 1c. **c, d,** More images scan from the other parts of the cross-section as blue line shows in **(a)**.

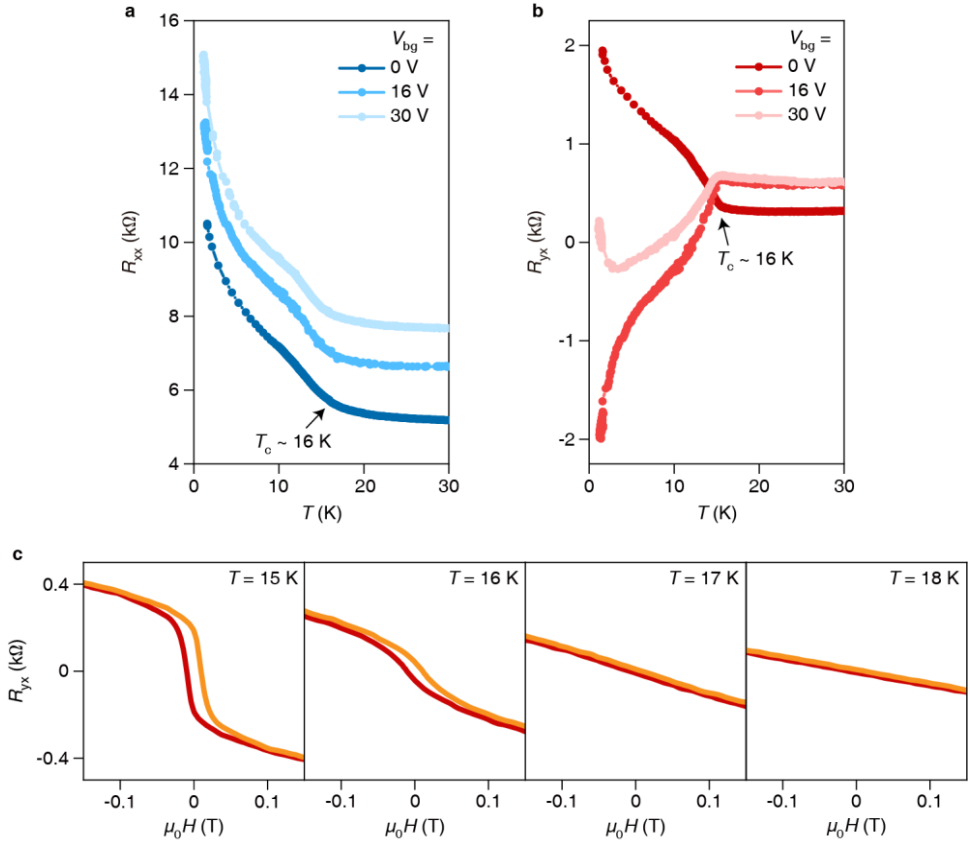

**Extended Data Fig. 2 FM-paramagnetic phase transition in monolayer device. a,b,** Temperature-dependent longitudinal **(a)** and Hall **(b)** resistance of the MnBi$_4$Te$_7$ monolayer device at different back gate voltages $V_{bg}$. The kink appears at $T_c \sim 16$ K. **c,** $R_{yx}$ - $\mu_0 H$ curves at different temperature. The hysteresis loop appears at temperature down to ~ 16 K.

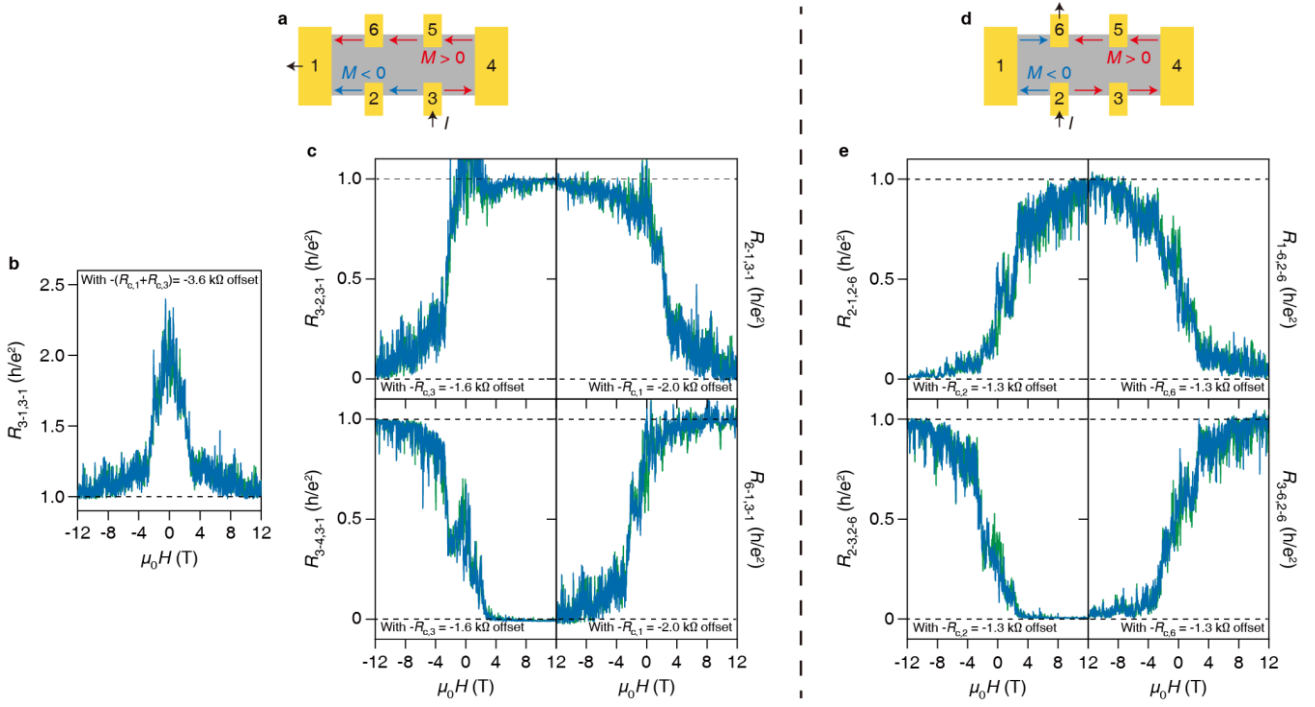

**Extended Data Fig. 3 Nonlocal measurements. a,** Schematic of the measurement configuration and chiral edge modes for panels **(b)** and **(c)**. The red and blue arrows indicate the chiral edge current for magnetization $M > 0$ and $M < 0$, respectively. **b, c,** Two- **(b)** and three-terminal **(c)** measurements with measurement configuration shown in **(a)**. **d,** Schematic of the measurement configuration and chiral edge modes for panels **(e)**. **e,** Three-terminal measurements with measurement configuration shown in **(d)**. $R_{\text{a-b,c-d}}$ represents that we apply current from lead $c$ to $d$ and measure voltage from lead $a$ to $b$. The results are subtracted with normal contact resistance with the values labeled in corresponding figures. These results are coincident with predictions from Landauer-Büttiker formula for chiral edge states.

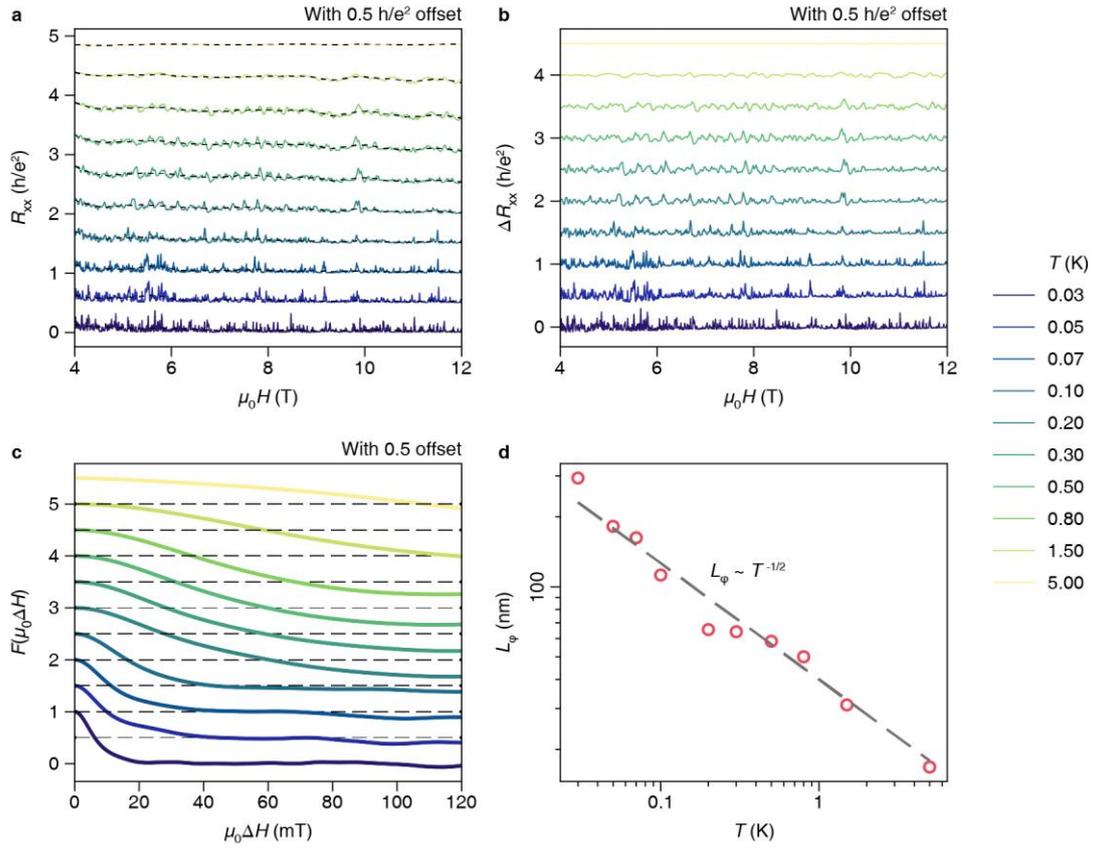

**Extended Data Fig. 4 Extract phase coherence length from UCFs a,** Continuous measurements of $R_{xx}$ v.s. $\mu_0 H$ at different $T$ in one thermal cycle. The dotted lines show the resistance background. **b,** $\Delta R_{xx}$ with resistance background subtracted to demonstrate UCFs. **c,** Autocorrelation function of $\Delta R_{xx}$ - $\mu_0 H$ curves at different $T$. **d,** Extracted phase coherence length $L_\varphi$ v.s. $T$. The calculation detail is shown in 'Method'.

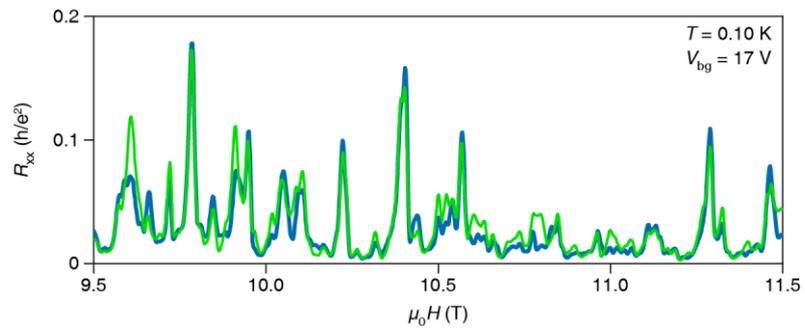

**Extended Data Fig. 5 Repeating measurements of $R_{xx}$ v.s. $\mu_0 H$ at $T$ = 0.10 K and $V_{bg}$ = 17 V.** The fluctuations are repeatable, confirming the existence of UCFs.

# Supplementary Information for

# Observation of topological Anderson Chern insulator phase

# in MnBi$_4$Te$_7$ monolayer

Anqi Wang[1,4,*], Bo Yin[1,4,*], Zikang Su[1,4,*], Shangjie Tian[7], Guoan Li[1,4], Xiaofan Shi[1,4], Xiao Deng[1,4], Yupeng Li[1], Zhiyuan Zhang[1,4], Xingchen Guo[1,4], Qinghua Zhang[1,8], Lin Gu[5], Xingjiang Zhou[1,4,9], Bingbing Tong[1], Peiling Li[1], Zhaozheng Lyu[1], Guangtong Liu[1,9], Fanming Qu[1,9], Ziwei Dou[1], Yuan Huang[3,†], Hechang Lei[2,6,†], Hongming Weng[1,4], Zhong Fang[1,4], Quansheng Wu[1,4,†], Li Lu[1,4,9,†], Jie Shen[1,9,†]

[1]Beijing National Laboratory for Condensed Matter Physics, Institute of Physics, Chinese Academy of Sciences, Beijing 100190, China

[2]School of Physics and Beijing Key Laboratory of Optoelectronic Functional Materials & Micro-Nano Devices, Renmin University of China, Beijing 100872, China

[3]School of Integrated Circuits and Electronics, Beijing Institute of Technology, Beijing 100081, China

[4]School of Physical Sciences, University of Chinese Academy of Sciences, Beijing 100049, China

[5]Beijing National Center for Electron Microscopy and Laboratory of Advanced Materials, Department of Materials Science and Engineering, Tsinghua University, Beijing 100084, China

[6]Key Laboratory of Quantum State Construction and Manipulation (Ministry of Education), Renmin University of China, Beijing 100872, China

[7]Anhui Key Laboratory of Magnetic Functional Materials and Devices, School of Materials Science and Engineering, Anhui University, Hefei 230601, China

[8]Yangtze River Delta Physics Research Center Co. Ltd, Liyang 213300, China

[9]Songshan Lake Materials Laboratory, Dongguan 523808, China

*These authors contributed equally to this work

†Corresponding author. Email: yhuang@bit.edu.cn (Y.H.), hlei@ruc.edu.cn (H.L.), quansheng.wu@iphy.ac.cn (Q.W.), lilu@iphy.ac.cn (L.L.), shenjie@iphy.ac.cn (J.S.)

# I. Symmetrization and anti-symmetrization procedures

We employ the standard symmetrization and anti-symmetrization procedures to separate the $R_{xx}$ and $R_{yx}$ components in the measurement. In magnetic field sweeping process, the symmetrization and anti-symmetrization procedures follow:

$$R_{xx}^{\uparrow}(\mu_0 H) = [R_{xx,raw}^{\uparrow}(\mu_0 H) + R_{xx,raw}^{\downarrow}(-\mu_0 H)]/2,$$
$$R_{xx}^{\downarrow}(\mu_0 H) = [R_{xx,raw}^{\downarrow}(\mu_0 H) + R_{xx,raw}^{\uparrow}(-\mu_0 H)]/2,$$
$$R_{yx}^{\uparrow}(\mu_0 H) = [R_{yx,raw}^{\uparrow}(\mu_0 H) - R_{yx,raw}^{\downarrow}(-\mu_0 H)]/2,$$
$$R_{yx}^{\downarrow}(\mu_0 H) = [R_{yx,raw}^{\downarrow}(\mu_0 H) - R_{yx,raw}^{\uparrow}(-\mu_0 H)]/2, \qquad (1)$$

where 'raw' in subscription means raw data, and '↑'/'↓' in superscript indicate the magnetic field sweeping direction.

Supplementary Figs. 1-4 demonstrate the $R_{xx}/R_{yx}$ - $\mu_0 H$ curves at $T$ = 0.10 K and different $V_{bg}$ before and after symmetrization/anti-symmetrization procedures.

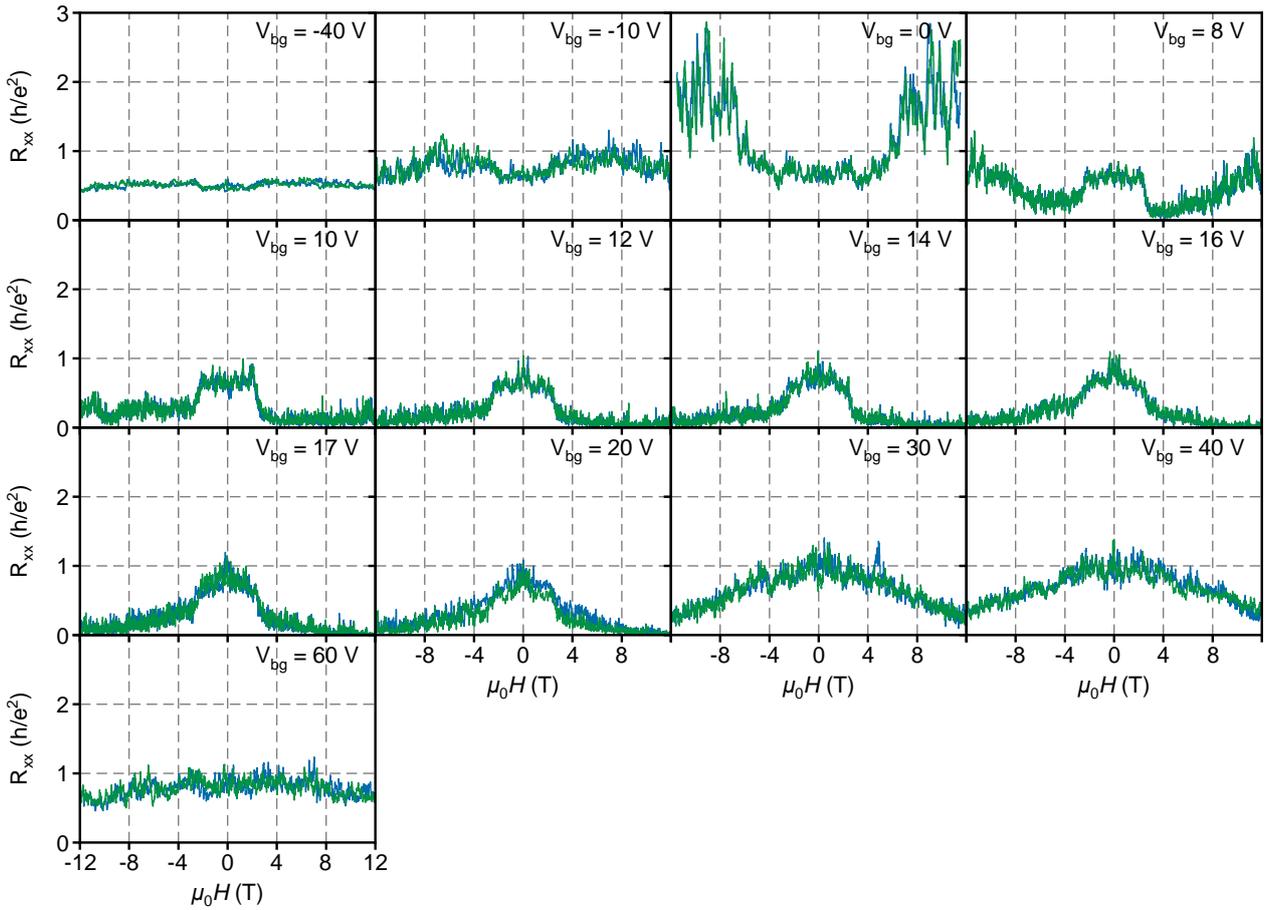

**Supplementary Fig. 1 $R_{xx}$ - $\mu_0 H$ curves at $T$ = 0.10 K and different $V_{bg}$ before symmetrization procedures.**

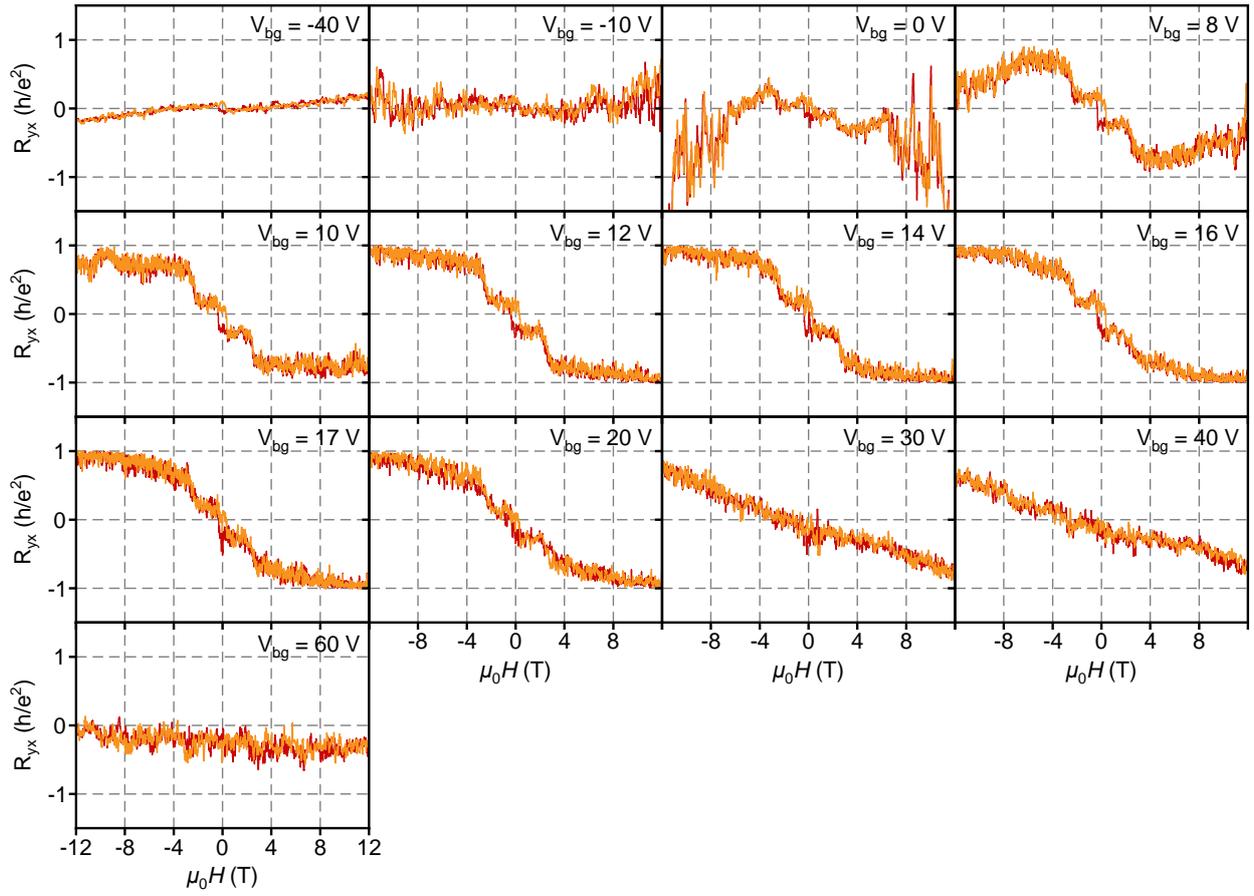

**Supplementary Fig. 2** $R_{yx}$ - $\mu_0 H$ curves at $T$ = 0.10 K and different $V_{bg}$ before anti-symmetrization procedures.

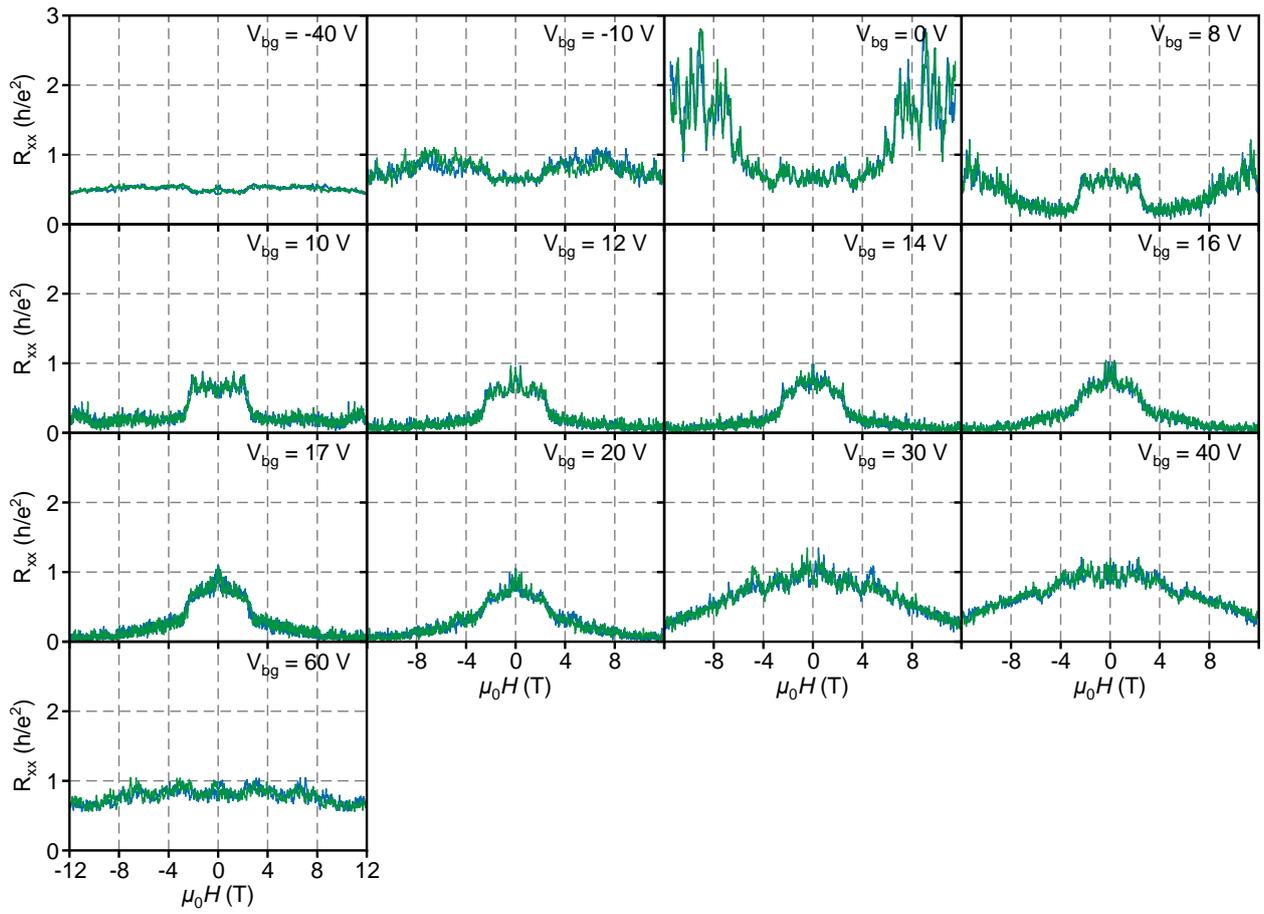

**Supplementary Fig. 3** $R_{xx}$ - $\mu_0 H$ curves at $T$ = 0.10 K and different $V_{bg}$ after symmetrization procedures.

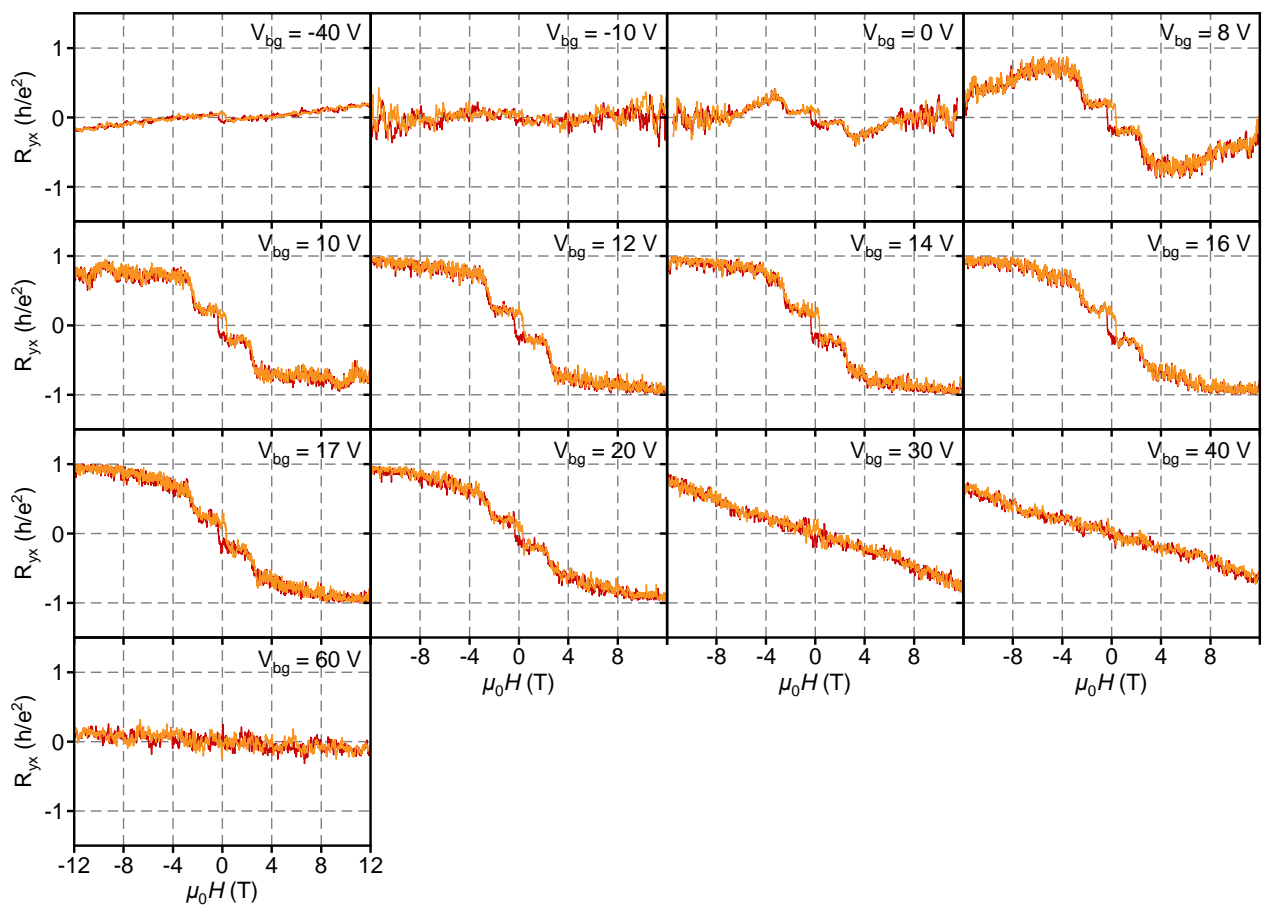

**Supplementary Fig. 4** $R_{yx}$ - $\mu_0 H$ curves at $T$ = 0.10 K and different $V_{bg}$ after anti-symmetrization procedures.

**II. Hysteretic behavior in gate sweeps**

The $R_{xx}/R_{yx}$ v.s. $V_{bg}$ curves exhibit hysteresis behavior due to the charging effect of the $Al_2O_3/SiO_2$ back gate dielectric layer. As shown in Supplementary Fig. 5a, when the gate voltage is swept in the same direction, the curves are almost coincident. However, when swept in opposite direction, a gate voltage shift of ~ 5 V is observed. To perform the symmetrization/anti-symmetrization procedures for $R_{xx}/R_{yx}$ - $V_{bg}$ curves, we ensure the gate voltages swept at $\mu_0 H = \pm 12$ T are in the same direction at same $T$, thereby avoiding the hysteresis effect. For symmetrized/anti-symmetrized $R_{xx}/R_{yx}$ - $V_{bg}$ curves at different $T$, we use the similar method as used in ref. 1 to correct the gate voltage hysteresis: by applying a ~ 5 V gate voltage shift in curves with different sweeping direction and make the $R_{xx}$ - $V_{bg}$ curves crossover at one point (the phase transition point $V_{bg}^T$). Supplementary Figure 5b demonstrates the original $R_{xx}/R_{yx}$ - $V_{bg}$ curves sweeping at $\mu_0 H = \pm 12$ T and different $T$ and Supplementary Fig. 5c gives the specific gate voltage shifts applied to each curve to correct the hysteretic behavior. The $R_{xx}/R_{yx}$ - $V_{bg}$ curves after gate voltage correction are shown in Fig. 2b.

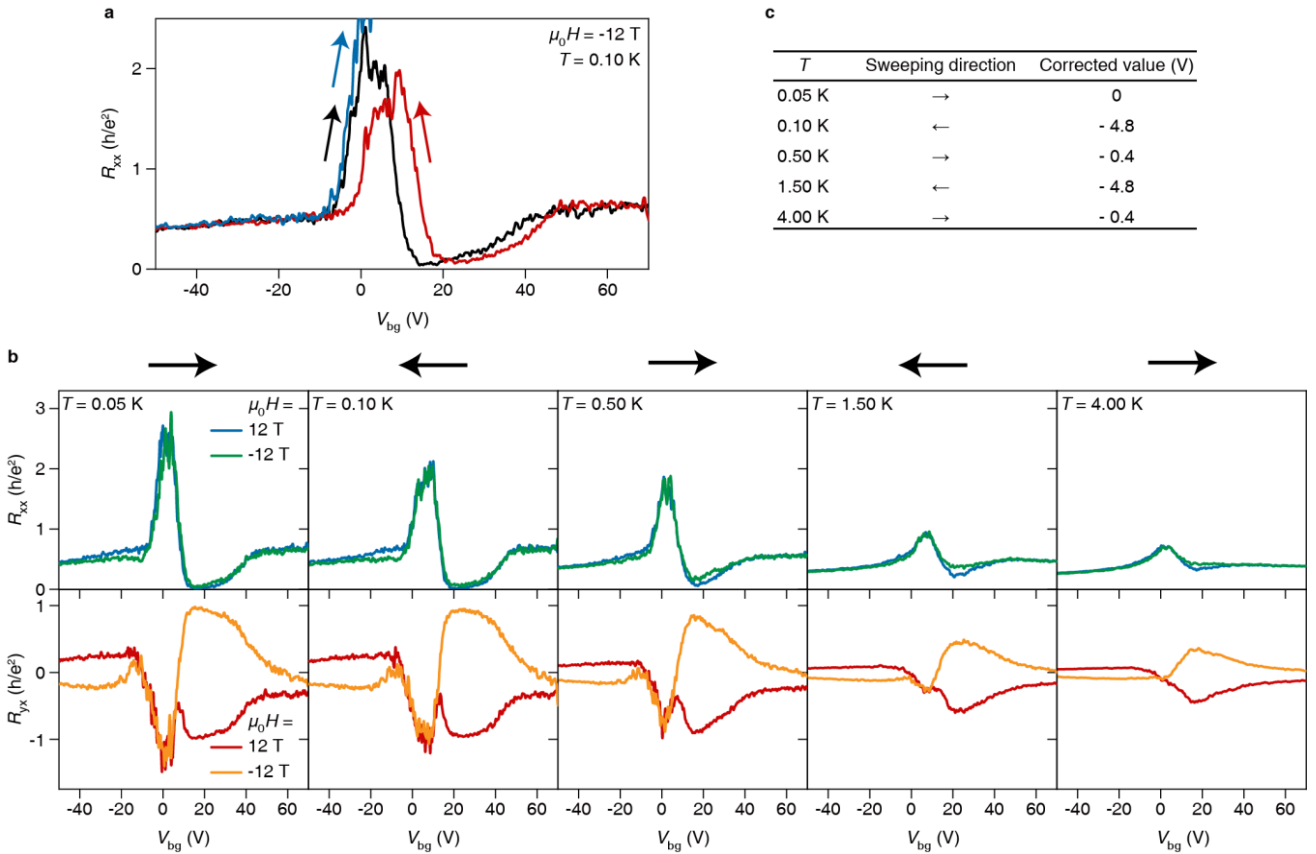

**Supplementary Fig. 5 Hysteretic behavior in gate sweeps. a,** Sweeping gate voltage at $\mu_0 H = -12$ T and $T = 0.10$ K with different directions. **b,** Original $R_{xx}/R_{yx}$ - $V_{bg}$ curves sweeping at $\mu_0 H = \pm 12$ T and different $T$ with the sweeping direction labeled as the black arrows on the top. **c,** The specific gate voltage shifts applied to each $R_{xx}/R_{yx}$ - $V_{bg}$ curve to correct the hysteretic behavior.